\documentclass[prl,twocolumn,superscriptaddress]{revtex4}
\usepackage{amssymb,amsmath,amsthm,graphicx}
\usepackage{latexsym}
\usepackage{bm}
\usepackage[]{hyperref}
\usepackage{cleveref}

\pdfoutput=1

\begin{document}

\title{Charting the AdS Islands of Stability with Multi-oscillators?}

\author{Matthew Choptuik}
\email{choptuik@physics.ubc.ca}
\affiliation{Department of Physics and Astronomy, University of British Columbia, 6224 Agricultural Road, Vancouver, B.C., V6T 1W9, Canada} 
\author{Jorge~E.~Santos}
\email{jss55@cam.ac.uk}
\affiliation{Department of Applied Mathematics and Theoretical Physics, University of Cambridge, Wilberforce Road, Cambridge CB3 0WA, UK} 
\author{Benson~Way}
\email{benson@phas.ubc.ca}
\affiliation{Department of Physics and Astronomy, University of British Columbia, 6224 Agricultural Road, Vancouver, B.C., V6T 1W9, Canada} 

\begin{abstract}
We propose the existence of an infinite-parameter family of solutions in AdS that oscillate on any number of non-commensurate frequencies.  Some of these solutions appear stable when perturbed, and we suggest that they can be used to map out the AdS ``islands of stability". By numerically constructing two-frequency solutions and exploring their parameter space, we find that both collapse and non-collapse are generic scenarios near AdS. Unlike other approaches, our results are valid on any timescale and do not rely on perturbation theory.
\end{abstract}

\maketitle

%\section{Introduction}
%---------------------------------------------------------------------------
{\bf~Introduction --} Unlike Minkowski and de-Sitter space where nonlinear stability has long been established \cite{friedrich86,Christodoulou:1993uv}, global anti-de Sitter (AdS) space with a reflecting boundary may allow arbitrarily small energy excitations to form black holes \cite{DafermosHolzegel2006,Dafermos2006}. The first evidence of this instability appeared in \cite{Bizon:2011gg}, 
where a set of initial data 
always seemed to lead to collapse,
even when the amplitude was arbitrarily small.

% MWC: Use some other term than non-collapsing?  Perhaps regular?
However, accumulated evidence \cite{Dias:2012tq,Maliborski:2013jca,Buchel:2013uba,Balasubramanian:2014cja,Bizon:2014bya,Balasubramanian:2015uua,Dimitrakopoulos:2015pwa,Green:2015dsa} suggests that there are families of initial data that do not form black holes.  While some such data are unstable and can produce black holes when perturbed, others do not seem to collapse under any small perturbation. Such stable non-collapsing data are said to lie within the \textit{islands of stability} \footnote{This term often refers to data that does not collapse within a certain timescale (typically scaling with the energy as $1/E$). For our purposes, we take this timescale to be infinite.}.  The extent of these islands remains an open problem, with important implications for the instability and its interpretation within the field theory dual provided by the AdS/CFT correspondence \cite{Maldacena:1997re,Gubser:1998bc,Witten:1998qj,Aharony:1999ti}. 

To address this question, let us begin with perturbation theory about $\mathrm{AdS}_d$. Consider a massless (real, or complex) scalar field $\varphi$, minimally coupled to gravity, within spherical symmetry. Let $t$ be the time measured by an observer at the boundary of AdS, and $x$ be a radial coordinate.  Take the perturbative expansion
\begin{equation}
\varphi(t,x)=\sum_{p=1}^{\infty} \epsilon^{p}\varphi^{(p)}(t,x)\;,\quad g_{\mu\nu}(t,x)=\sum_{p=0}^{\infty} \epsilon^{p}g_{\mu\nu}^{(p)}(t,x)\;,
\end{equation}
where $g^{(0)}_{\mu\nu}$ is the metric for $\mathrm{AdS}_d$. 

At linear order, the perturbative solution consists of a linear combination of normal modes
\begin{equation}
%\varphi^{(1)}=\sum_{n=0}^{\infty}a^{\pm}_n\varphi^{\pm}_n\;,\qquad \varphi^{\pm}_n(t,x)=e^{\pm i\omega^{(0)}_nt}P_n(x)\;,
\varphi^{(1)}(t,x)=\sum_{n=0}^{\infty}(a^{+}_ne^{+i\omega^{(0)}_nt}+a^{-}_ne^{- i\omega^{(0)}_nt})P_n(x)\;,
\end{equation}
where $\omega^{(0)}_n=d-1+2n$, the $P_n$ form a set of orthogonal functions,
and $a^{\pm}_n$ are complex coeffcients.
If $\varphi$ is real, then we must also have $\frac{1}{2}a^-_n=\frac{1}{2}(a^+_n)^*$.

The terms at higher order depend upon the choice of $a^{\pm}_n$ at linear order. If $a^{\pm}_n\neq0$ for at least two distinct $n$, a secular term proportional to $\epsilon^3 t$ appears in the scalar field at third order, leading to an apparent breakdown of perturbation theory at $t\sim1/\epsilon^2$. Much evidence \cite{Bizon:2011gg,Buchel:2012uh,Buchel:2013uba,Balasubramanian:2014cja,Bizon:2014bya,Buchel:2014xwa,Balasubramanian:2015uua,Bizon:2015pfa,Green:2015dsa,Freivogel:2015wib,Choptuik:2017cyd} has found that gravitational collapse, when it occurs, happens on this timescale of $t\sim1/\epsilon^2$.  

However, now consider allowing $a^{\pm}_n\neq0$ for only one value of $n$. In this case, the secular terms that appear at higher order can be removed using a standard Poincar\'e-Lindstedt resummation procedure \cite{Bizon:2011gg}. It is then possible to construct a scalar field configuration that oscillates at a single frequency.  Such solutions are often called \textit{oscillons} \cite{Maliborski:2013jca,Fodor:2015eia} if $\varphi$ is a real scalar field, and \textit{boson stars} \cite{Dias:2011at,Liebling:2012fv,Buchel:2013uba,Choptuik:2017cyd} for configurations of a complex scalar field where the metric does not oscillate. Away from spherical symmetry, similar solutions exist in pure gravity called \emph{geons} \cite{Dias:2011ss,Horowitz:2014hja,Martinon:2017uyo}.  For now, we use the generic term \textit{oscillator}. Oscillators can be roughly viewed as nonperturbative extensions of normal modes.

Oscillators are non-collapsing and the ones near AdS appear stable.  To date, all initial data within the islands of stability appear to be ``close'' to an oscillator.  But can the extent of the islands be determined more precisely? Note that data ``close'' to a normal mode still allows for any number of nonzero $a^{\pm}_n$, so naive perturbation theory appears to break down at $t\sim1/\epsilon^2$ for stable data as well.

Attempts to resum perturbation theory has led to some suggestive results (see e.g.\cite{Balasubramanian:2014cja,Craps:2014vaa,Craps:2014jwa,Buchel:2014xwa,Bizon:2015pfa,Evnin:2015wyi,Yang:2015jha,Bizon:2013xha,Dimitrakopoulos:2015pwa,Green:2015dsa}).  In particular, the two-timescale formalism (TTF), introduced and developed in \cite{Balasubramanian:2014cja,Craps:2014vaa,Craps:2014jwa,Buchel:2014xwa}, allows a slow time dependence in the amplitudes $a^\pm_n(\epsilon^2t)$.  The TTF equations contain a scaling symmetry and extra conserved quantities not present in the full equations. This scaling symmetry implies that non-collapsing solutions within the TTF at finite amplitude can be extended towards zero amplitude \cite{Dimitrakopoulos:2015pwa}. These results, together with numerical evidence, suggest that both collapse and non-collapse are generic in the sense that both kinds of data have finite measure in the $\epsilon\to0$ limit.

Here, we take a different, non-perturbative approach. We will construct solutions to the full nonlinear equations that do not collapse on any timescale.  By exploring the parameter space of these solutions, we hope to chart the extent of these islands, particularly near AdS. 

To explain our approach, let us recall the construction of an oscillator.  First, AdS is linearly perturbed to obtain a normal mode with frequency $\omega^{(0)}_n$, for some particular $n$. Oscillators can be generated by correcting the normal mode with higher orders in perturbation theory, where $\omega^{(0)}_n$ also receives perturbative corrections. Nonperturbatively, oscillators have a spectral expansion
\begin{equation}
\varphi(t,x)=\sum_{k=-\infty}^{\infty}\sum_{l=0}^{\infty}A_{k,l}e^{ik \omega_1 t}P_l(x)\;,
\end{equation}
where the period $\omega_1$ can be used as a parameter, recovering AdS when $\omega_1=\omega^{(0)}_n$.  Each choice of $n$ generates a one-parameter family of oscillators.  These solutions can be obtained numerically by treating $t$ as periodic and solving a boundary value problem.

Now let us attempt to repeat this process again. Consider perturbing the oscillator:
\begin{equation}
\varphi(t,x)=\bigg(\sum_{k,l}A_{k,l}e^{ik \omega_1 t}P_l(x)\bigg)+\epsilon e^{-i\omega_2 t}\delta\varphi(t,x)\;,
\end{equation}
where $\delta\varphi(t,x)$ is periodic in time with period $\omega_1$, and the metric is similarly perturbed. Note that the function $\varphi(t,x)$ is not periodic, since $\omega_1$ and $\omega_2$ need not be commensurate frequencies.  However, at linear order in $\epsilon$,  $\omega_2$ will appear as an eigenvalue in the equations of motion, and all perturbation functions will remain periodic in time with period $\omega_1$.  This system can therefore be solved as a boundary value problem using standard methods for finding eigenvalues. 

Just as in the case for AdS itself, the reflecting boundary of AdS leads to a spectrum of normal modes for the frequency $\omega_2$. We can use one of these normal modes to generate a new set of \textit{double-oscillators}. Non-perturbatively, such a solution takes the spectral form
\begin{equation}\label{doublespec1}
\varphi(t,x)=\sum_{k_1,k_2=-\infty}^{\infty}\sum_{l=0}^{\infty}A_{k_1,k_2,l}e^{ik_1 \omega_1 t+ik_2 \omega_2 t}P_l(x)\;,
\end{equation}
which is `periodic' on two frequencies $\omega_1$ and $\omega_2$, forming a two-parameter family.

We propose the following method of constructing such solutions \eqref{doublespec1}. Consider the alternative spectral expansion
\begin{equation}\label{doublespec2}
\varphi(t_1,t_2,x)=\sum_{k_1,k_2,l}A_{k_1,k_2,l}e^{ik_1 \omega_1 t_1+ik_2 \omega_2 t_2}P_l(x)\;,
\end{equation}
which is periodic on $t_1$ and $t_2$. When the equations of motion are in first-order form, time-dependence only appears in the derivative $\partial_t$. A comparison between \eqref{doublespec1} and \eqref{doublespec2} suggests that we should replace $\partial_t\rightarrow \partial_{t_1}+\partial_{t_2}$.  The double-oscillator can then be obtained by solving a boundary value problem in coordinates $t_1$, $t_2$, and $x$, where $t_1$, and $t_2$ are periodic with period $\omega_1$ and $\omega_2$, respectively.   $t_1$ and $t_2$ can be viewed as coordinates on a torus around which the line $t=t_1+t_2$ wraps.  If the periods $\omega_1$ and $\omega_2$ are non-commensurate, then the line parametrised by $t$ will be dense on this torus.  

One can continue this process ad-infinitum, generating \textit{multi-oscillators} that are periodic on more and more frequencies. This creates an infinite-parameter family of solutions, all of which are non-collapsing.  The apparent nonlinear stability of single-oscillators suggests that nearby multi-oscillators are also stable.

These multi-oscillators bear some resemblance to quasi-periodic solutions found within the TTF \cite{Balasubramanian:2014cja,Green:2015dsa}, where each normal mode is assumed to have a different periodic behaviour $a_n(\epsilon^2t)=\alpha_ne^{-i\beta_n t}$.  We note, however, that quasi-periodic solutions within the TTF form a two-parameter family, while we have an infinite-parameter family. We also mention that quasi-periodic behaviour has also been seen within dynamical evolution \cite{Maliborski:2013jca,Choptuik:2017cyd,Biasi:2017kkn}. We speculate that some of these may have a representation as a multi-oscillator. 

Though we have explained the construction of multi-oscilators for a scalar field in spherical symmetry, the arguments for their existence and method of construction apply equally well with fewer spatial symmetries, with other field configurations like pure gravity, and even situations without AdS boundary conditions. So long as the perturbation of oscillators continues to yield normal modes, additional frequencies can be added.  

We wish to construct such solutions and explore their parameter space.  We were successful in constructing double-oscillators for a real scalar, but such solutions depend on three coordinates, as shown in \eqref{doublespec2}. Here, we instead present results using a complex scalar field where the first frequency dependence $e^{i\omega_1t_1}$ factors out, reducing the problem to only two coordinates. This type of solution may be viewed as a boson star in the frequency $\omega_1$, but an oscillon in the frequency $\omega_2$.

\textbf{Numerical Construction --} We set the AdS length scale and gravitational constant to unity, and the spacetime dimension $d$ to $5$. Our metric ansatz is
\begin{subequations}\label{phipidef}
\begin{align}
\mathrm ds^2&=\frac{1}{\cos^2 x}\left(-\alpha \beta^2\mathrm dt^2+\frac{dx^2}{\alpha}+\sin^2x\,\mathrm d\Omega_{3}\right)\;,\\
\varphi&=\cos^4x\,e^{i\omega_1 t}(\varphi_++i\varphi_-)\;,\label{scalarredef}
%\varphi&=\cos^4x\,e^{i\omega_1 t}(\varphi_r+i\varphi_i)\;,
\end{align}
\end{subequations}
where we have defined
\begin{equation}\label{metricredef}
\alpha=1-\sin^2x\cos^{4}x\,A\;,\qquad \beta=1-\cos^8x\,\delta\;,
\end{equation}
and $A$, $\delta$, and $\varphi_\pm$ are real functions of $t$ and $x$. We introduce first-order variables $\Phi_+$, $\Phi_-$, $\Pi_+$, $\Pi_-$ via
\begin{subequations}\label{phipidef}
\begin{align}
\cos x\,&\partial_x\varphi_\pm-4\sin x\varphi_\pm=\Phi_\pm\;,\\
&\partial_t\varphi_\pm=\alpha\beta\frac{\Pi_\pm}{\cos x}\pm\omega_1\varphi_\pm\;.
%\cos x\,&\partial_x\varphi_r-4\sin x\varphi_r=\Phi_r\;,\\
%\cos x\,&\partial_x\varphi_i-4\sin x\varphi_i=\Phi_i\;,\\
%&\partial_t\varphi_r=\alpha\beta\frac{\Pi_r}{\cos x}+\omega_1\varphi_i\;,\\
%&\partial_t\varphi_i=\alpha\beta\frac{\Pi_i}{\cos x}-\omega_1\varphi_r\;.
\end{align}
\end{subequations}
The Hamiltonian constraint takes the form
\begin{equation}\label{Hconstraint}
\partial_t A=2\frac{\cos^4 x}{\sin x}\,\alpha^2\beta(\Phi_+\Pi_++\Phi_-\Pi_-)\;,
\end{equation}
while the remaining equations of motion are
\begin{subequations}
\begin{align}
\cos x\,&\partial_x\delta-\sin x(8+\cos^8x\,S)\delta=-\sin x\,S\\
\sin x\,&\partial_xA+\cos x(4+\sin^2x\cos^8x\,S)A=\cos^3 x\,S\\
\partial_t\Phi_\pm&=\beta(\alpha\partial_x\Pi_\pm-A_\Phi\tan x\,\Pi_\pm)\pm\omega_1\Phi_\mp\;,\\
\partial_t\Pi_\pm&=\beta(\alpha\partial_x\Phi_\pm+A_\Pi\cot x\,\Phi_\pm)\pm\omega_1\Pi_\mp\;,
%\end{align}
%\end{subequations}
%\begin{subequations}
%\begin{align}
%\partial_t\Phi_r&=\beta(\alpha\partial_x\Pi_r-A_\Phi\tan x\,\Pi_r)+\omega_1\Phi_i\;,\\
%\partial_t\Phi_i&=\beta(\alpha\partial_x\Pi_i-A_\Phi\tan x\,\Pi_i)-\omega_1\Phi_r\;,\\
%\partial_t\Pi_r&=\beta(\alpha\partial_x\Phi_r+A_\Pi\cot x\,\Phi_r)+\omega_1\Pi_i\;,\\
%\partial_t\Pi_i&=\beta(\alpha\partial_x\Phi_i+A_\Pi\cot x\,\Phi_i)-\omega_1\Pi_r\;,
\end{align}
\end{subequations}
where
\begin{subequations}
\begin{align}
S&=\Phi_+^2+\Phi_-^2+\Pi_+^2+\Pi_-^2\;,\\
A_\Phi&=3-\frac{1}{2}[9-5\cos(2x)]\cos^4x A\;,\\
A_\Pi&=3-\frac{1}{2}[3-5\cos(2x)]\sin^2x\cos^2x A\;.
\end{align}
\end{subequations}
In what follows, the Hamiltonian constraint \eqref{Hconstraint} as well as the definitions \eqref{phipidef} are not solved directly, and are instead used as a check of numerics. 

To find the boson stars and their perturbations, set
\begin{subequations}
\begin{align}
\Phi_+(t,x)&=\Phi_0(x)+\epsilon[\cos (\omega_2 t)\delta\Phi_+(x)]\;,\\
\Phi_-(t,x)&=\epsilon[\sin (\omega_2 t)\delta\Phi_-(x)]\;,\\
\Pi_+(t,x)&=\epsilon[\sin (\omega_2 t)\delta\Pi_+(x)]\;,\\
\Pi_-(t,x)&=\Pi_0(x)+\epsilon[\cos (\omega_2 t)\delta\Pi_-(x)]\;,\\
A(t,x)&=A_0(x)+\epsilon[\cos (\omega_2 t)\delta\delta(x)]\;,\\
\delta(t,x)&=\delta_0(x)+\epsilon[\cos (\omega_2 t)\delta A(x)]\;,\\
\varphi_+(t,x)&=\varphi_0(x)+\epsilon[\cos (\omega_2 t)\delta\varphi_+(x)]\;,\\
\varphi_-(t,x)&=\epsilon[\sin (\omega_2 t)\delta\varphi_-(x)]\;,
\end{align}
\end{subequations}
where we have chosen a specific phase in time.  Setting $\epsilon=0$ will yield a set of ordinary differential equations (ODEs) that can be solved to obtain the boson star. These boson stars are parametrised by the frequency $\omega_1$. A linear expansion in $\epsilon$ will yield another set of ODEs for the perturbation functions in the form of an eigenvalue problem with $\omega_2$ as an eigenvalue. We solve these using Fourier spectral methods. We use a quarter Fourier grid in the coordinate $x$ that exploits the symmetries of the functions. By the redefinitions \eqref{scalarredef} and \eqref{metricredef}, these symmetries plus finiteness of the functions ensure that all required boundary conditions are satisfied. 

Once the boson stars and their perturbations have been computed, we can use them to find double-oscillators. Because the $\omega_1$ periodicity is not manifest in the equations, we can find the desired solutions by treating $t$ as a periodic coordinate with period $\omega_2$. The symmetries in the functions along $t$ allow us to use a half Fourier grid. The full equations of motion are solved as a boundary value problem, using a Newton-Raphson method with the perturbed boson stars as initial estimates.

\begin{figure}
\centering
\includegraphics[width=.4\textwidth]{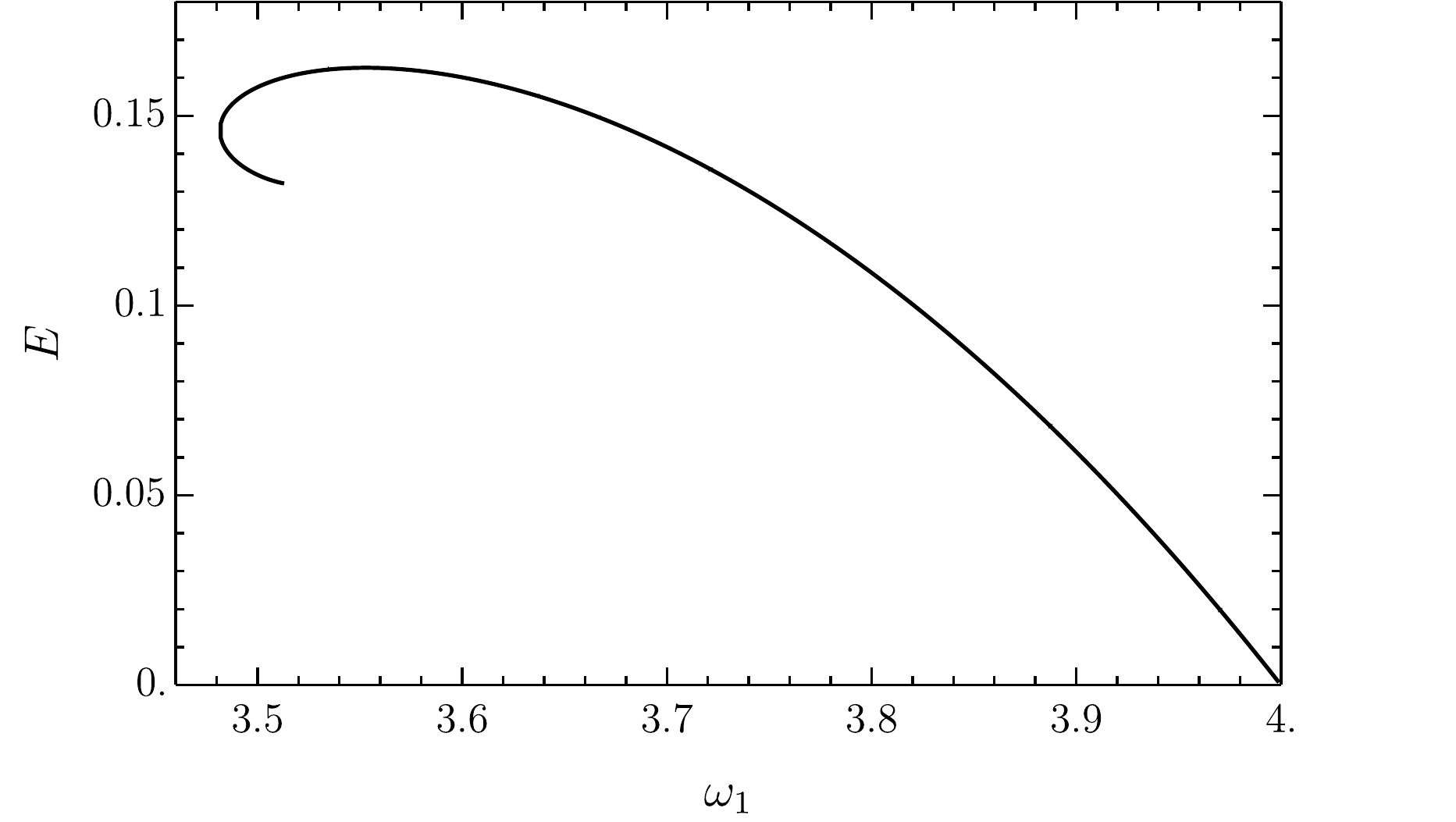}
\caption{Energy of boson stars versus their frequency.}\label{fig:ebosonstar}
\end{figure}

\textbf{Results --}  In Fig. \ref{fig:ebosonstar}, we show the energy of the lowest frequency branch of boson stars as a function of the frequency $\omega_1$. This curve is typical of oscillators, and has been produced elsewhere \cite{Dias:2011at,Liebling:2012fv,Buchel:2013uba,Choptuik:2017cyd}.  The point corresponding to $\omega_1=4$ is pure AdS, where the frequency merely represents a perturbative normal mode. Naturally, there are other branches of boson stars that are generated from other normal modes of AdS, but we do not consider them here.

This figure shows a turning point at $\omega_1\approx 3.55$, where the energy is maximal. Such turning points are typically associated with a change in linear stability, and is a common feature in many gravitational systems \cite{poincare:1885,Sorkin:1981jc,Sorkin:1982ut,Arcioni:2004ww}.  

The dynamical stability of boson stars was studied previously in \cite{Buchel:2013uba,Choptuik:2017cyd}, where indeed solutions on the AdS side of the turning point 
appeared to be stable, while solutions on the other side are unstable.  The endpoint of unstable boson star evolution 
depends on how the star is  perturbed: one generically sees either collapse to a black hole or migration towards some oscillating solution \cite{Choptuik:2017cyd}. 

\begin{figure}
\centering
\includegraphics[width=.4\textwidth]{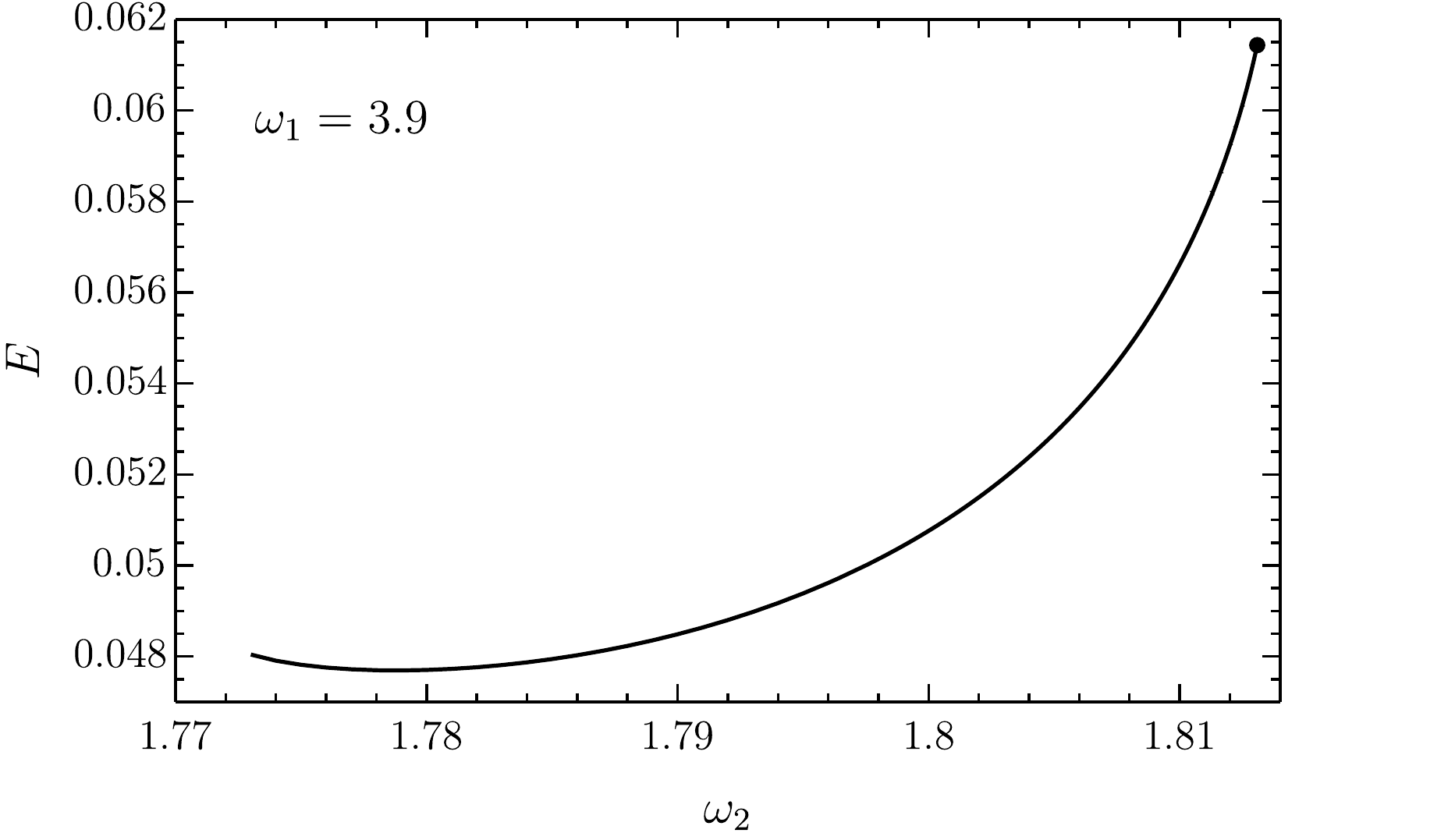}
\includegraphics[width=.4\textwidth]{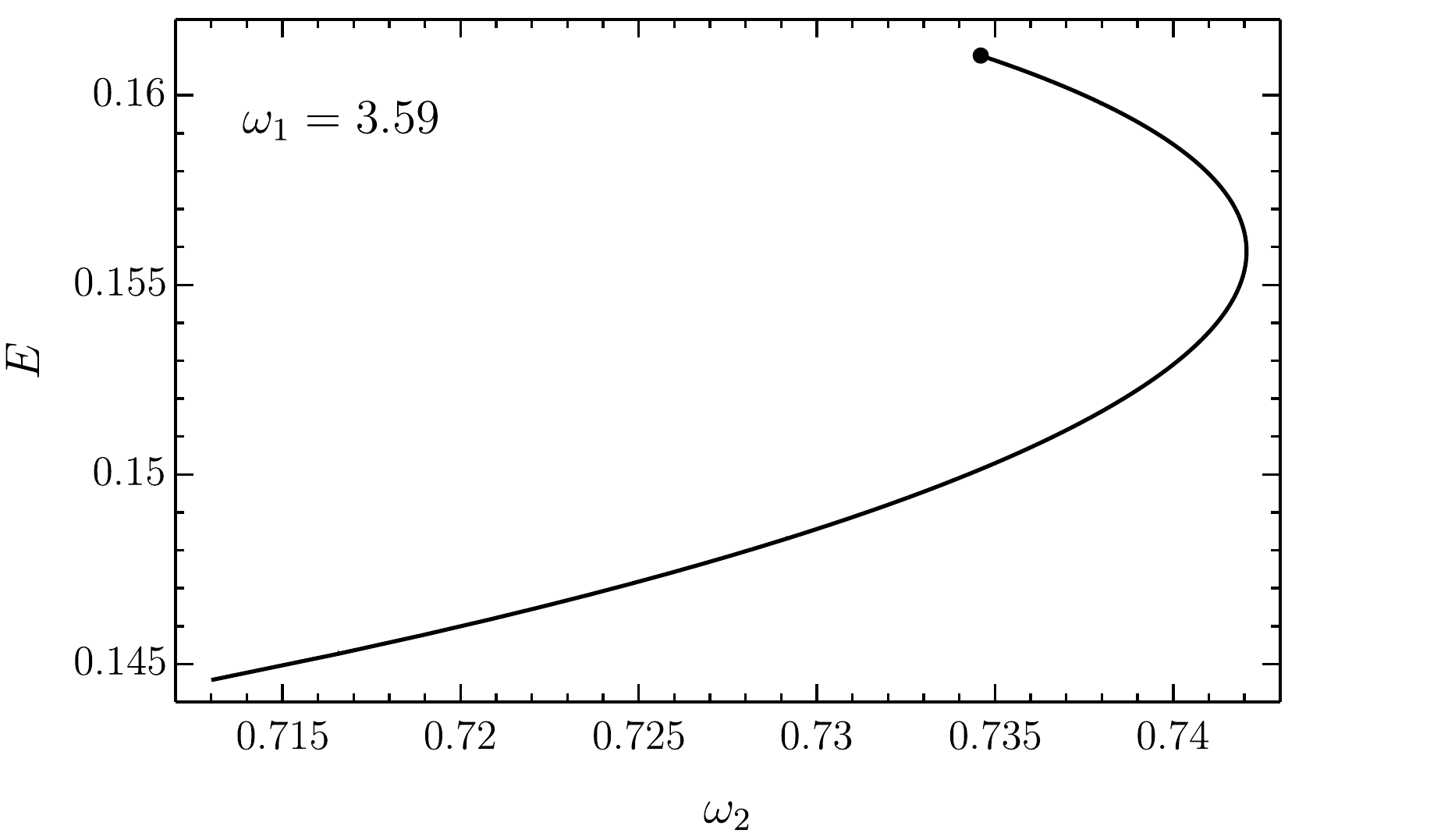}
\caption{Energy of double-oscillators at fixed $\omega_1$. The boson star solutions are located at the black dots.}\label{fig:eosc}
\end{figure}

In Fig. \ref{fig:eosc}, we fix the frequency $\omega_1$ and plot the energy of double-oscillators as a function of $\omega_2$. In the top panel, $\omega_1=3.9$ and in the bottom panel, $\omega_1=3.59$. The boson star coincides with the dot in each figure, where the frequency represents a perturbative normal mode.  Unlike the boson stars where moving away from AdS increases the energy, these double-oscillators decrease in energy as one moves away from the boson star. %At present, we do not have an explanation that accounts for this behaviour. 

The top panel in Fig. \ref{fig:eosc} has a turning point in the energy, where the energy is minimal. We expect that the lower panel has such a turning point as well, but were unable to reach it with our limited computational resources.  By analogy with the pure boson star and similar situations \cite{Harada:1998ge,Uryu:2000dw,Shibata:2001zj,Friedman:2001pf,Arcioni:2004ww,Lahiri:2007ae,Boshkayev:2012bq,Haensel:2016pjp}, this turning point may come with a change in dynamical stability. Physically, the expectation is that linear energy fluctuations correspond to changes in frequency.  But near a turning point, the energy does not change to first order, suggesting that some frequency becomes a zero mode and is thus unstable on one side of the turning point.  We emphasise, however, that it remains unclear whether these stability arguments apply to our current situation. We leave more detailed questions of stability and dynamical evolution to future work.

In the top panel of Fig. \ref{fig:eosc}, $\omega_2$ decreases as the double-oscillators move away from the boson star, while in the bottom panel $\omega_2$ initially increases before decreasing again. That is, the double oscillators in the bottom panel have a turning point in frequency. Such a turning point occurs for double-oscillators with $\omega_1\lesssim3.6$.

\begin{figure}
\centering
\includegraphics[width=.4\textwidth]{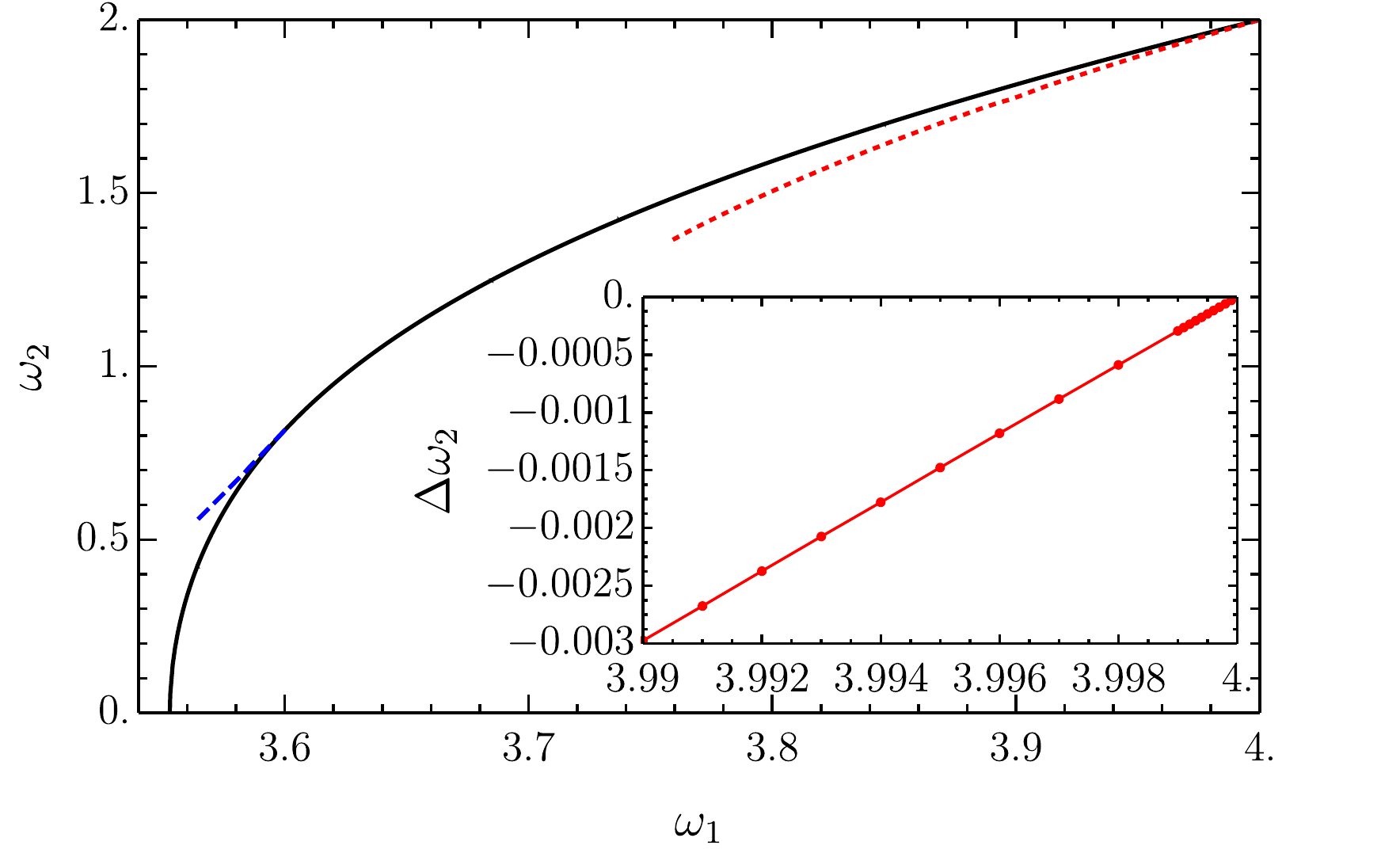}
\caption{Parameter space of double-oscillators.  The solid line are the boson stars, the dashed blue curve is where double-oscillators have a turning point in the frequency, and the dotted red line is where oscillators have a turning point in the energy. The 
inset zooms in near AdS, and plots the frequency difference, $\Delta\omega_2$,
relative to the 
boson star}\label{fig:w12}
\end{figure}  
We are now in a position to present the space of double-oscillator solutions parametrised by $\omega_1$ and $\omega_2$. In Fig. \ref{fig:w12}, the black solid line shows the lowest perturbative frequency $\omega_2$ of the boson star.  The point $\omega_1=4$, $\omega_2=2$ corresponds to pure AdS, with both of these frequencies representing AdS normal modes.  At $\omega_1\approx 3.55$,  corresponding to the turning point seen in Fig. \ref{fig:ebosonstar}, $\omega_2$ becomes a zero mode.   For $\omega_1\lesssim 3.55$, the normal mode frequency of the boson star $\omega_2$ becomes purely imaginary, corresponding to an unstable mode.  There are therefore no double-oscillators generated by this mode for $\omega_1\lesssim 3.55$. 

The dotted red curve locates the turning points in energy, such as the one that occurs in the top panel of Fig.~\ref{fig:eosc}. If these turning points mark a change in stability, we expect that oscillators above this red line to be stable, and those below it to be unstable. 

The dashed blue curve locates the turning points in frequency $\omega_2$, such as the one that occurs in the bottom panel of Fig. \ref{fig:eosc}.  Double-oscillators at these values of $\omega_1$ exist below the blue curve, and the region immediately below this curve contains two oscillator solutions.

Now we examine the space of double oscillators in the neighbourhood of AdS. The inset of Fig. \ref{fig:w12} shows the difference in frequency $\Delta\omega_2$ between the double-oscillator and the boson star normal mode in the region near AdS. Note that the line does not become tangent to $\Delta\omega_2=0$ near $\omega_1=4$, suggesting an open set of non-collapsing data in the neighbourhood of AdS. This is in agreement with arguments in \cite{Dimitrakopoulos:2015pwa}. Note that unlike \cite{Dimitrakopoulos:2015pwa}, our results are constructive (rather than merely proving existence), non-perturbative, and valid on any timescale. 

\begin{figure}
\centering
\includegraphics[width=.4\textwidth]{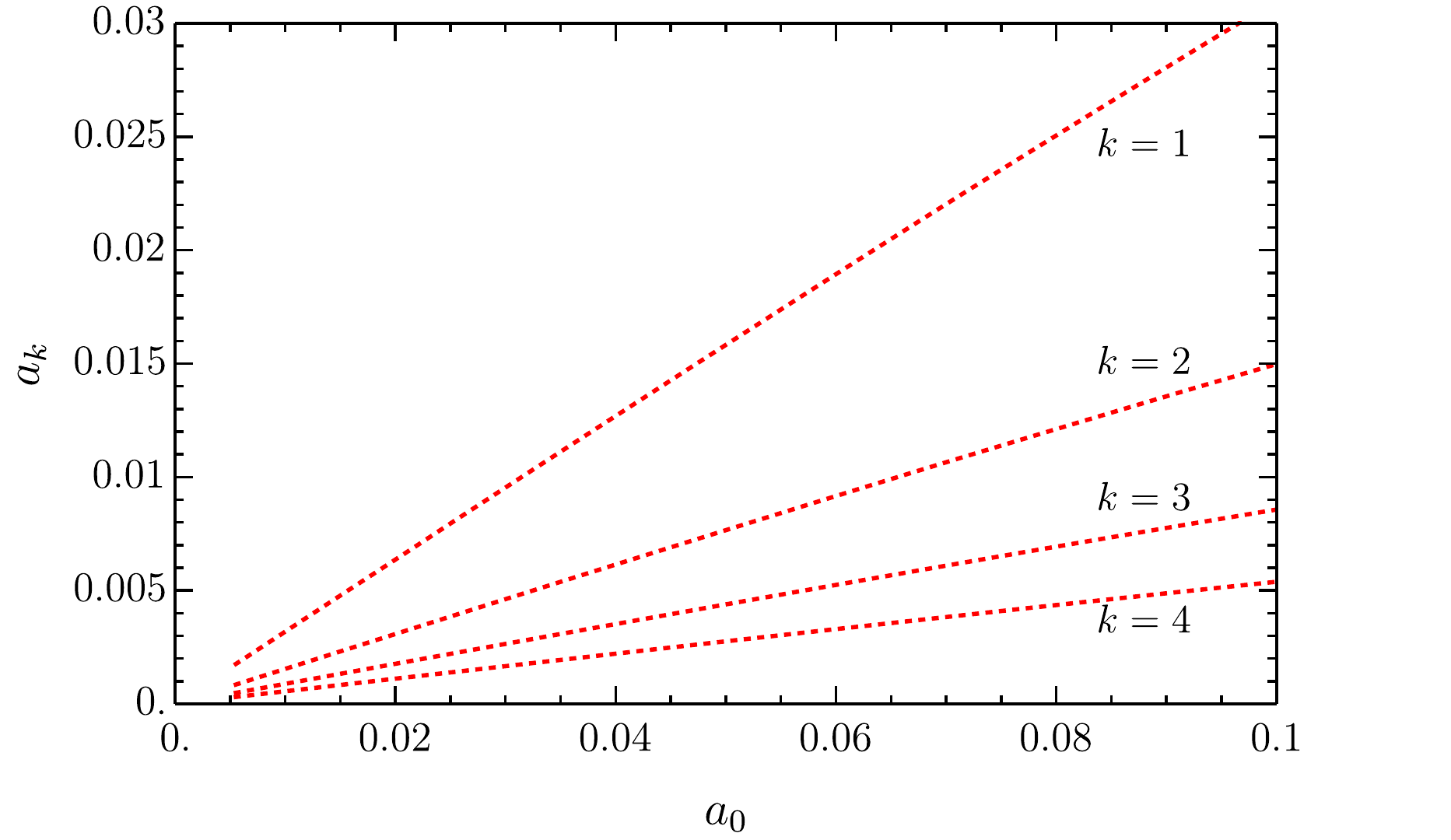}
\caption{Scalar field of double-oscillators at the energy turning point, evaluated at $t=0$ (where $\varphi$ is real) and projected onto the orthogonal functions $\varphi(t=0,x)=\sum a_kP_k(x)$.}\label{fig:jacobi}
\end{figure}  

The solutions at the energy turning point can be projected onto the orthogonal functions $P_k(x)$, as we show in Fig. \ref{fig:jacobi}.  We observe that these lines have nonzero slope near AdS, which is another indication of an open set of non-collapsing data \cite{Dimitrakopoulos:2015pwa}. We also note that the modes do not reach equal-amplitude, which agrees with numerical evidence suggesting that equal-amplitude, two-mode initial data will eventually form black holes \cite{Bizon:2011gg,Balasubramanian:2014cja,Buchel:2014xwa,Green:2015dsa,Bizon:2015pfa,Deppe:2015qsa}.

\textbf{Discussion --}  
We have proposed the existence of a family of non-collapsing solutions that oscillate on any number of frequencies, and provided a method of constructing them that does not rely on any perturbative approximation nor dynamical evolution.  

This infinite-parameter family is unusually large. In a sense, the entire space of initial data can also be viewed as an infinite-parameter family.  The existence of so many multi-oscillators may account for the apparent stability of boson stars and oscillons. It is also tempting to conjecture that the entire island of stability lies within the multi-oscillator family. 

We have constructed and mapped out part of a two-frequency section of this family.  We found that double-oscillators contain a line of turning points in their energy.  If such turning points indeed mark a change in stability, then they also mark a boundary to the islands of stability. One could presumably map out more of the islands of stability by searching for turning points among additional frequencies.  But, like charting any coastline, the entire boundary of these islands of stability cannot be determined by any finite process. 

%------------------------------------------------------------
{\bf~Acknowledgements --} We thank Oscar Dias, Stephen Green, Gary Horowitz, and Luis Lehner for helpful comments, and for reading an earlier version of the manuscript. J.E.S. is supported in part by STFC grants PHY-1504541 and ST/P000681/1.   M.W.C. and B.W. are supported by NSERC.
\bibliography{refs}{}
\end{document}